\documentclass[floatfix,twocolumn,aps,prl,showpacs,amsmath,amssymb]{revtex4}
\usepackage[latin1]{inputenc}

\usepackage[dvips]{graphicx}

\usepackage{dcolumn}
\usepackage{bm}
\usepackage{color}
\usepackage{ulem} 
\usepackage{url}
\usepackage[colorlinks=true ,citecolor=blue]{hyperref}
\usepackage{amsmath,amssymb}

\begin{document}

\title{ Dynamical Mass Generation in Pseudo Quantum Electrodynamics with Four-Fermion Interactions}

\author{Van S\'ergio Alves$^{1}$, Reginaldo O. C. Junior$^{2}$, E. C. Marino$^2$, and Leandro O. Nascimento$^{3,4}$ }
\affiliation{$^1$ Faculdade de F\'\i sica, Universidade Federal do Par\'a, Av.~Augusto Correa 01, Bel\'em PA, 66075-110, Brazil  \\
$^2$   Instituto de F\'\i sica, Universidade Federal do Rio de Janeiro, C.P. 68528, Rio de Janeiro RJ, 21941-972, Brazil \\
$^3$ International Institute of Physics, Campus Universit\' ario Lagoa Nova, C.P. 1613, Natal RN, 59078-970, Brazil  \\
$^4$ Faculdade de Ci\^encias Naturais, Universidade Federal do Par\'a, C.P. 68800-000, Breves, PA,  Brazil  }

\date{\today}

\begin{abstract}

We describe dynamical symmetry breaking in a system of massless Dirac fermions with both electromagnetic and four-fermion interactions in  (2+1) dimensions. The former is described by the Pseudo Quantum Electrodynamics (PQED) and the latter is given by the so-called Gross-Neveu action. We apply the Hubbard-Stratonovich transformation and the large$-N_f$ expansion in our model to obtain a Yukawa action. Thereafter, the presence of a symmetry broken phase is inferred from the non-perturbative Schwinger-Dyson equation for the electron propagator. This is the physical solution whenever the fine-structure constant is larger than a critical value $\alpha_c(D N_f)$. In particular, we obtain the critical coupling constant $\alpha_c\approx 0.36$ for $D N_f=8$., where $D=2,4$ corresponds to the SU(2) and SU(4) cases, respectively, and $N_f$ is the flavor number. Our results show a decreasing of the critical coupling constant in comparison with the case of pure electromagnetic interaction, thus  yielding a more favorable scenario for the occurrence of dynamical symmetry breaking. For two-dimensional materials,in application in condensed matter systems, it implies an energy gap at the Dirac points or valleys of the honeycomb lattice.
\begin{center}
Subject Areas: Condensed Matter Physics, Graphene, Field Theory Methods
\end{center}

\end{abstract}


\maketitle

\section{\textbf{I.\,Introduction}}

Two-dimensional quantum field theories are relevant for describing the electronic interactions in thin materials, for instance, graphene and transition metal dichalcogenides \cite{review}. They also have been applied as minimal models for describing some particular features of Quantum Chromodynamics, in particular, the quark confinement \cite{Roberts}. For condensed matter physics, we may comment on the discovery of topological states of matter, which are described by topological order instead of spontaneous symmetry breaking. \cite{Thouless,Haldane}. A key ingredient for describing these states is the opening and closing of an energy gap at the Dirac points, changing the topology of the system and breaking some discrete symmetry. Hence, the investigation about dynamical mass generation in these theories still a very important branch of physics \cite{Bernevig}. 

Having in mind two-dimensional materials, we may separate the interactions into two classes: one due to the electric charge and the other due to microscopic interactions that emerge into these systems, for instance, phonons, impurities, and  disorder. The first case is described by PQED, whose derivation has been made in Ref.~\cite{marino}. The main procedure is to confine the matter current into the plane, but to keep photons free to propagate out of this plane. One of the advantages is to recover the usual Coulomb interaction in the plane instead of the logarithmic potential, typical of Quantum Electrodynamics in (2+1)-dimensions (QED3). Several features of PQED have been discussed in literature \cite{PRX2015,VLWJF,CSBTemperature,Teber,marino1,unitarity,Kovner,marinorubens}. In the static limit, the Coulomb interaction renormalizes the Fermi velocity \cite{Vozmediano} of electrons in graphene, accordingly to the experimental measurements in Ref.~\cite{Elias}. At the neutrality point in graphene, Lorentz symmetry is expected to be recovered. The second kind of interactions are not described by an unique model. In general grounds, it is expected to be given by some electron-electron action.

For graphene, in the absence of external magnetic field and considering the full electromagnetic interaction, it has been shown the emergence of a quantized valley Hall conductivity at low temperatures. Essentially, this effect is explained by the dynamical generation of mass in the electronic spectrum, above a certain critical coupling constant $\alpha_c$ \cite{PRX2015}. The role of PQED for describing topological states of matter has been discussed in Ref.~\cite{BJP}. In particular, for massive Dirac systems, it  was shown the existence of an emerging quantum Hall effect \cite{ArxivQHE}. It is worth to remember that the Hall conductivity is only defined in the SU(2) representation, where the mass breaks time-reversal symmetry. In the SU(4) representation, dynamical mass generation has been calculated for PQED in both zero \cite{VLWJF} and finite \cite{CSBTemperature} temperatures, describing chiral symmetry breaking. Similar results have been obtained for QED3 \cite{Appelquist1985,Maris2} and QED4 earlier, see Refs.~\cite{Roberts} for a detailed review about these studies. On the other hand, the description of dynamical symmetry breaking in gauge theories with additional four-fermion interactions has been less discussed \cite{Miransky}. In particular, this case for PQED has not been investigated until now.

In this paper, we calculate the pattern of dynamical mass generation in PQED, including a Gross-Neveu interaction \cite{GN} at zero temperature, both SU(2) and SU(4) representations are discussed. The main goal is to understand how the critical behavior of the theory from the competition between the two coupling constants, namely, the fine-structure constant $\alpha$ and the dimensional constant $G$ that emerges from the four-fermion interaction. Furthermore, we argue that this model may be applied for describing interactions in two-dimensional systems. Indeed, PQED captures the electromagnetic interaction and the four-fermion term effectively describes some microscopic interaction, typical of these materials. This seems be the case, because after we include an auxiliary field into the theory, we obtain a Yukawa action, which has been shown to describe electron-phonon interaction at low-energies, see Ref.~\cite{Saito}. It is shown that the presence of microscopic interaction, which may be effectively described by the Yukawa action, does not cancel the quantum valley Hall effect, because the dynamically generated masses are still quantized, leading to a quantized valley current, see Ref.~\cite{PRX2015}.

The outline of this paper is the following: in Sec.~II we introduce our model within the PQED approach and the usual Gross-Neveu model. In Sec.~III we apply the Hubbard-Stratonovich transformation  in the Gross-Neveu model, using the large $N_{f}$ expansion. In Sec.~IV we calculate the auxiliary and Gauge-field propagators in the  large-$N_{f}$ expansion. In Sec.~V we write the Schwinger-Dyson equation for the matter field. In Sec.~VI we calculate the mass function $\Sigma(p)$ in the unquenched-rainbow approach. In Sec.~VII we calculate the physical mass. In Sec.~VIII we summarize and outlook our main results. We also include three appendixes, where we derive the integral equation to wavefunction renormalization (see Appendix A), derive the differential equation for the mass function (see Appendix B), and perform numerical testes (see Appendix C).

\section{\textbf{II. The Model}}
 We assume that the Dirac electrons will interact through the electromagnetic interaction, which in 2D is described by  PQED \cite{marino}. Furthermore, we assume a finite four-fermion interaction. The corresponding Lagrangian reads
\begin{equation}
{\cal L}=\frac{1}{4} F_{\mu \nu}\left[\frac{2}{\sqrt{-\Box}}\right] F^{\mu\nu} + i\bar\psi_a\partial\!\!\!/ \psi_a+j^{\mu}\, A_{\mu}-\frac{G}{2}(\bar\psi_a \psi_a)^2
\label{action}
\end{equation}
where, $j^\mu=e\,\bar\psi_a\gamma^\mu\psi_a$ is the matter current. $\psi_a$ is a four-component Dirac field, $\bar\psi_a =\psi_a^\dagger\gamma^0$ is its adjoint, $F_{\mu \nu}$ is the usual field intensity tensor of the U(1) Gauge field $A_\mu$, which intermediates the electromagnetic interaction in 2D (pseudo electromagnetic field),  $\gamma^\mu$ are rank-2 Dirac matrices, and $a=1,...,N_f$ is a flavor index. The coupling constant $e^2= 4\pi\alpha$ is conveniently written in terms of $\alpha$, the fine-structure constant in natural units. $G$ is the coupling constant related to the four-fermion interaction, i.e, the Gross-Neveu interaction. The dimension of $G$ is the inverse of energy, namely, $[G]=[M]^{-1}$. Without four-fermion interactions ($G=0$), an SU(4) version of this model has been recently used to study dynamical gap generation and chiral symmetry breaking in graphene \cite{VLWJF}. For $\alpha=0$, Eq.~(\ref{action}) is the Gross-Neveu model. 

The bare Gauge-field propagator, in the Landau Gauge, reads
\begin{equation}
G_{0,\mu\nu}(p)=\frac{P_{\mu\nu}}{2\sqrt{p^2}},  \label{photonbare}
\end{equation}
where $P_{\mu\nu}=\delta_{\mu\nu}-p_\mu p_\nu/p^2$ is the transversal operator. The bare fermion propagator is
\begin{equation}
S_{0,F}(p)=\frac{1}{\gamma^\mu p_\mu}.
\end{equation}

Next, we introduce an auxiliary field in order to transform the action from the four-fermion interaction to a Yukawa action. This transformation is also known as the Hubbard-Stratonovich transformation.

\section{ \textbf{III. Auxiliary Field}}

Let us first consider the fermionic terms in Eq.~(\ref{action}), i.e, the Gross-Neveu action ${\cal L}_{\rm{GN}}$, given by
\begin{equation}
{\cal L}_{\rm{GN}}=i\bar\psi_a\partial\!\!\!/ \psi_a-\frac{G}{2}(\bar\psi_a \psi_a)^2. \label{actionGN}
\end{equation}

In the large $N_f$-expansion, we define a new coupling constant $g= G N_f$, such that $g$ is meant to be fixed for $N_f$ large. Next, we introduce an auxiliary field $\varphi(x)$ in Eq.~(\ref{actionGN}) in order to obtain a three-linear vertex interaction. Therefore, we perform the following transformation
\begin{equation}
{\cal L}_{\rm{GN}}\rightarrow {\cal L}_{\rm{GN}}+\frac{1}{2g}\left(\varphi-\frac{g}{\sqrt{N_f}}\bar\psi_a\psi_a\right)^2, \label{transf}
\end{equation}
which gives the transformed action
\begin{equation}
{\cal L}_{GN}=  i\bar\psi_a\partial\!\!\!/ \psi_a-\frac{\varphi}{\sqrt{N_f}} \bar\psi_a\psi_a+\frac{\varphi^2}{2g}. 
\label{action2}
\end{equation}

It is straightforward, from the minimal principle for the auxiliary field $\delta{\cal L}/\delta\varphi=0$, to show that
\begin{equation}
\varphi=\frac{g}{\sqrt{N_f}}\bar\psi_a\psi_a,
\end{equation}
proving that the transformation in Eq.~(\ref{transf}) does not change the dynamics  of the Gross-Neveu model in Eq.~(\ref{actionGN}) (See Eq.~(\ref{transf})). Here, it is more convenient to consider Eq.~(\ref{action2}). The propagator of the auxiliary field is given by
\begin{equation}
\Delta_{0,\varphi}=\frac{1}{1/g},
\end{equation}
which clearly has no dynamics at bare level. Within this approach, Eq.~(\ref{action}) becomes
\begin{eqnarray}
{\cal L}&=& \frac{1}{4} F_{\mu \nu}\left[\frac{2}{\sqrt{-\Box}}\right] F^{\mu\nu} + i\bar\psi_a\partial\!\!\!/ \psi_a+j_a^{\mu}\, A_{\mu} \nonumber \\
&-&\frac{\varphi}{\sqrt{N_f}} \bar\psi_a\psi_a+\frac{\varphi^2}{2g}.
\label{actionend}
\end{eqnarray}

Before, we use the Schwinger-Dyson equation for the electron propagator, we need to obtain the propagators for the Gauge and scalar field. In the next section, we apply the large$-N_f$ expansion in order to obtain these full propagators.

\section{ \textbf{IV. The Large-$N_f$ expansion for both $\varphi$ and $A_\mu$}}

The Schwinger-Dyson equation for the auxiliary field $\varphi(x)$ is
\begin{equation}
\Delta^{-1}_\varphi=\Delta^{-1}_{0,\varphi}-\Pi(p),
\end{equation}
where $\Delta^{-1}_\varphi$ is the full propagator of $\varphi$ and $\Pi(p)$ is given by the fermionic loop, given by
\begin{equation}
\Pi(p)=-{\rm Tr} \int\frac{d^3k}{(2\pi)^3} S_F(p-k) S_F(k),
\end{equation} 
where $S_F(p)$ is the full electron propagator. Here, we use the bare electron propagator in order to find an analytical result for $\Delta_\varphi$. Hence $\Pi(p)=\sqrt{p^2} x_0$, with $x_0=D/16$ ($D=4$ is the rank of the Dirac matrices), and
\begin{equation}
\Delta_\varphi(p)=\frac{1}{1/g+\sqrt{p^2}  x_0}.
\end{equation}
Remarkably, the self-energy of the auxiliary field yields a nontrivial dynamics for the propagator of the auxiliary field.

The Schwinger-Dyson equation for the Gauge field $A_\mu$ is
\begin{equation}
G^{-1}_{\mu\nu}=G^{-1}_{0,\mu\nu}-\Pi_{\mu\nu}, \label{SDphoton}
\end{equation}
where $G_{\mu\nu}$ is the full propagator of the Gauge field and $\Pi_{\mu\nu}$ is the vacuum polarization tensor \cite{Luscher}. Eq.~\ref{SDphoton} yields
\begin{equation}
G_{\mu\nu}=G_{0,\mu\alpha}[\delta^{\alpha}_\nu-\Pi^{\alpha\beta}G_{0,\beta\nu}]^{-1}. \label{solphoton}
\end{equation}
The vacuum polarization tensor may be decomposed into
\begin{equation}
\Pi_{\mu\nu}=e^2 \Pi_1 P_{\mu\nu}+e^2\Pi_2\epsilon_{\mu\nu\alpha}p^\alpha,  \label{Pidec}
\end{equation}
where $\Pi_1$ and $\Pi_2$ are known functions \cite{Luscher}. Using Eq.~\ref{photonbare} and Eq.~\ref{Pidec} in Eq.~\ref{solphoton}, we find
\begin{equation}
G_{\mu\nu}=\Delta_1 P_{\mu\nu}+ \Delta_2 \epsilon_{\mu\nu\alpha}p^\alpha,
\end{equation}
where
\begin{equation}
\Delta_1=\frac{2 \sqrt{p^2}-e^2 \Pi_1}{(2\sqrt{p^2}-e^2 \Pi_1)^2+e^4p^2 \Pi^2_2}
\end{equation}
and
\begin{equation}
\Delta_2=\frac{e^2\Pi_2}{(2\sqrt{p^2}-e^2 \Pi_1)^2+e^4p^2 \Pi^2_2}.
\end{equation}

From the trace properties of the Dirac matrices, it is possible to show that the term $\Delta_2$ does not contribute for dynamical symmetry breaking. Indeed, one may keep this term in the Gauge-field propagator and the result shall be the same at the end of the calculations. Physically, this is because this term breaks time-reversal symmetry from the very beginning. In the SU(4) case, we have $\Delta_2=0$. Assuming a small electric charge and $\Pi_1=-p N_f D/32$, we have \cite{VLWJF}
\begin{equation}
G_{\mu\nu}(p)=\frac{P_{\mu\nu}}{\sqrt{p^2}\left(2+\frac{\lambda D}{ 32}\right)},
\end{equation}
where $\lambda=e^2 N_f$ is the coupling constant in the large-$N_f$ expansion. $D=2$ yields the result in the SU(2) representation, which the Dirac field has two components. $D=4$ yields the result in the SU(4) representation, which the Dirac field has four components.

\section{ \textbf{V. \,Schwinger-Dyson Equation of the Matter field}}

\begin{figure}[htb]
\centering
\includegraphics[scale=0.4]{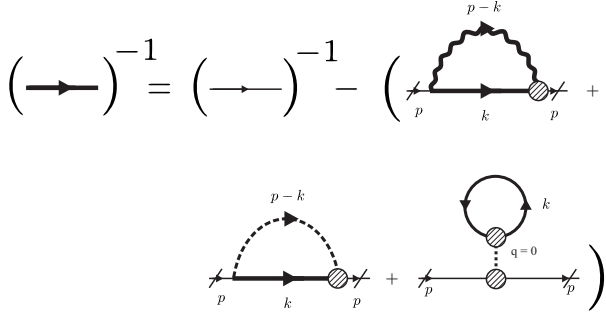}
\caption{(Color online) Diagrammatic representation of Eq.~(\ref{sd}). The bold lines are the full electron propagator and the thin line is the free electron propagator. The dashed line is the auxiliary-field propagator and the wave-like line is the Gauge-field propagator, both of them in the large-$N_f$ expansion. The cross hatched circle is the full vertex. Note that we have neglected all corrections to the vertex functions, which is called rainbow approach.} \label{figdia}
\end{figure}

The Schwinger-Dyson equation for the electron reads \cite{Roberts}
\begin{equation}
S^{-1}_F(p)=S_{0F}^{-1}(p)-\Xi (p),
\label{sd}
\end{equation}
where $S_{0F}$ and $S_{F}$ are, the free- and interacting-electron propagators, respectively. $\Xi(p)$ is the electron self-energy, which has two contributions $\Xi(p)=\Xi^{\alpha}(p)+\Xi^g(p)$, where $\Xi^\alpha(p)$ and $\Xi^g(p)$ are the electron self-energies due to the electromagnetic and four-fermion interactions, respectively. The diagrammatic representation is shown in Fig.~\ref{figdia}. The electron-self energies are given by
\begin{eqnarray}\label{sea}
\Xi^\alpha (p)=\frac{\lambda}{N_f} \int\frac{d^3k}{(2\pi)^3} \gamma^{\mu}S_F(k)\gamma^{\nu}\, G_{\mu\nu}(p-k)\,
\end{eqnarray}
and
\begin{eqnarray}\label{seg}
\Xi^g (p)=\frac{1}{N_f} \int\frac{d^3k}{(2\pi)^3} \Delta_\varphi(p-k)S_F(k)+ \nonumber
\\ - \Delta_\varphi(0) {\rm Tr}\int\frac{d^3k}{(2\pi)^3} S_F(k).
\end{eqnarray}

In order to obtain analytical solutions of the Schwinger-Dyson equation, it is convenient to rewrite the full fermion propagator as \cite{Maris2}
\begin{equation}
S_F^{-1}(p)=p_{\mu}\gamma^{\mu}\,A(p)-\Sigma (p), \label{full1}
\end{equation}
where $A(p)$ is usually called the wavefunction renormalization, and $\Sigma(p)$ the mass function. Inserting Eq.~(\ref{full1}) into Eq.~(\ref{sd}), we obtain the integral equation
\begin{eqnarray}
\label{sigmap}
\Sigma(p)&=&\frac{2\lambda}{N_f}\int\frac{d^3k}{(2\pi)^3}\,\frac{\Sigma(k)\,}{A^{2}(k)k^2+\Sigma^2(k)}\,\frac{1}{\sqrt{(p-k)^2}(2+\frac{\lambda D}{32})} \nonumber \\
&+& \frac{1}{N_f} \int\frac{d^3k}{(2\pi)^3} \frac{\Delta_\varphi (p-k)\Sigma(k)}{A^{2}(k) k^2+\Sigma^2(k)}  \nonumber \\ 
&-&\Delta_\varphi(0)\int\frac{d^3k}{(2\pi)^3} \frac{{\rm Tr} \Sigma(k) }{A^{2}(k) k^2+\Sigma^2(k)}.
\end{eqnarray}

From now on, we shall consider $A(p)\approx 1$, see Appendix A and Appendix C for more details.

\section{\textbf{VI. The Dynamical Solution for $\Sigma(p)$}}

The third term on the rhs of Eq.~(\ref{sigmap}) does not change momentum. Hence, we focus on the dynamical solutions of the mass function driven by a kernel with $p$-dependence. This regime is obtained when both Gauge and auxiliary propagators change momentum with the electron propagator.  
\begin{eqnarray}
\label{sigmap3}
\Sigma(p)=\frac{2\lambda}{N_f}\int\frac{d^3k}{(2\pi)^3}\,\frac{\Sigma(k)\,}{k^2+\Sigma^2(k)}\,\frac{1}{\sqrt{(p-k)^2}(2+\frac{\lambda D}{32})} \nonumber \\
+ \frac{1}{N_f} \int\frac{d^3k}{(2\pi)^3} \frac{\Sigma(k)}{k^2+\Sigma^2(k)} \frac{1}{1/g+\sqrt{(p-k)^2}  x_0}.
\end{eqnarray}

We use spherical coordinates $d^3k=k^2 dk \sin \theta d\theta d\phi$, hence, the polar integral gives a factor $2\pi$. Next, we solve the angular integral in both the first and second terms in the rhs of Eq.~(\ref{sigmap3}). 

Let us first consider only the integral which is proportional do $\lambda$, i.e, the first term of the rhs of Eq.~(\ref{sigmap3}). By defining $u\equiv p^2+k^2-2pk \cos\theta$ and changing the integral variable into $u$, we find
\begin{equation}
\frac{2\lambda}{4 \pi^2 N_f} \frac{1}{(2+\frac{\lambda D}{32})} \int_0^\infty \frac{k^2 dk \Sigma(k)}{k^2+\Sigma^2(k)}\left(\frac{|p+k|-|p-k|}{pk}\right).
\end{equation}
For the second term in the rhs, the same procedure yields
\begin{eqnarray}
&&\frac{1}{4 \pi^2 x_0 N_f p} \int_0^\infty \frac{k dk \Sigma(k)}{k^2+\Sigma^2(k)} \{ [|p+k|-|p-k|\nonumber\\
&-&\frac{1}{x_0 g} \ln \left[\frac{(x_0 g)^{-1}+|p-k|}{(x_0 g)^{-1}+|p+k|}\right]\}.
\end{eqnarray}
Therefore, the integral equation becomes
\begin{eqnarray}
&\Sigma(p)&=\frac{C_2}{p}\int_0^\infty\frac{k dk \Sigma(k)}{k^2+\Sigma^2(k)} \ln \left[\frac{(x_0 g)^{-1}+|p-k|}{(x_0 g)^{-1}+|p+k|}\right]\nonumber\\
&&+\frac{C_1}{p}\int_0^\infty \frac{k dk \Sigma(k)}{k^2+\Sigma^2(k)}\left(|p+k|-|p-k|\right), \label{eqint0}
\end{eqnarray}
where
\begin{equation}
C_1=\frac{2\lambda}{4\pi^2N_f(2+\lambda D/32)}+\frac{g}{N_f (g x_0) 4\pi^2},
\end{equation}
and
\begin{equation}
C_2=-\frac{g}{N_f (g x_0)^2 4\pi^2}.
\end{equation}

At this level, it is convenient to write a scale-invariant integral equation (without any dimensional parameter). By defining $\Sigma(p,g)\equiv f(pg)/g$, $x\equiv g p$, and $y\equiv g k$, we find 
\begin{eqnarray}
&f(x)&=\frac{g C_2}{x}\int_0^\infty\frac{y dy f(y)}{y^2+f^2(y)} \ln \left[\frac{(x_0)^{-1}+|x-y|}{(x_0)^{-1}+|x+y|}\right]\nonumber\\
&&+\frac{C_1}{x}\int_0^\infty \frac{y dy f(y)}{y^2+f^2(y)}\left(|x+y|-|x-y|\right). \label{eqint1}
\end{eqnarray}

From Eq.~(\ref{eqint1}), we have the functional dependence of the mass function on the momentum and other parameters: $\Sigma(p)=g^{-1}f(pg, \lambda,N_f)$. This result shows that the critical values for dynamical mass generation are related either for $\lambda$ or $N_f$. The dimensional coupling constant $g$ only changes the scale of the external momentum $p$. This result is exact. Indeed, with $g=0$, PQED with massless Dirac fermion is scale invariant and has a critical coupling constant $\lambda_c$ or a critical number of flavors $N_c$ \cite{VLWJF}. The breaking of scale invariance due to the coupling constant $g\neq 0$ may not change these values within the continuum approximation.

We may obtain analytical solutions by introducing an ultraviolet cutoff $\Lambda$, thus converting Eq.~(\ref{eqint1}) into a differential equation at high-external-momentum regime, namely, $x \gg f(x)$ (see Appendix A). In this case, we have
\begin{equation}
\frac{d}{dx}\left(x^2\frac{df(x)}{dx}\right)+\frac{N_c}{4N_f}f(x)=0, \label{eulereq}
\end{equation}
where
\begin{equation}
N_c(\lambda)= \frac{2}{\pi^2}\left(\frac{2\lambda}{2+\lambda D/32}+\frac{16}{D}\right)  \label{Nc}
\end{equation}
is the critical number of flavors. The mass function is nontrivial only if $N_f\leq N_c$ (similar to QED3 with $G=0$ \cite{Appelquist1985}).
This may be derived from the conditions
\begin{equation}\label{uvsup}
\lim_{x\rightarrow g\Lambda}\left(x \,\frac{df(x)}{dx}+f(x)\right)=0,
\end{equation}
and
\begin{equation}\label{irsup}
\lim_{x\rightarrow 0}x^2\,\frac{df(x)}{dx}=0,
\end{equation}
representing the ultraviolet (UV) and infrared (IR) asymptotic conditions, respectively.

The solutions of Euler's differential equation are
\begin{equation}
f(x)=A_+ x^{a_{+}}+ A_- x^{a_{-}},
\label{EulerSol}
\end{equation}
where $a_{\pm}=-1/2\pm1/2\sqrt{1-N_c/N_f}$ and $A_+$ and $A_-$ are arbitrary constants.
The real part of the solution  is
\begin{equation}
Re\{f(x)\}=\frac{F}{\sqrt{x}} \cos[\gamma \ln x+\theta_0],  \label{Rewritten}
\end{equation}
where $F$ and $\theta_0$ are arbitrary real constants. Indeed, it is straightforward to check that it satisfies Euler's differential equation. Using Eq.~(\ref{Rewritten}) in Eq.~(\ref{eulereq}), we find that the real constant $\gamma$ is given by 
\begin{equation}
\gamma=\frac{1}{2}\sqrt{\frac{N_c}{N_f}-1}. \label{gamma}
\end{equation}
The mass function reads $\Sigma(p)=g^{-1}f(gp)$. Eq.~(\ref{Rewritten}) with the critical point in Eq.~(\ref{Nc}) is our desired analytical solution. We shall use this to calculate the generated mass.

Next, we compare this criticality with some known results for QED3 \cite{Appelquist1985} and PQED \cite{VLWJF} without four-fermion interaction. From Eq.~(\ref{Nc}) and $\lambda=4\pi\alpha N_f$, we have 
\begin{equation}
N_c(\alpha)=\frac{2}{\pi^2}\left(\frac{8\pi\alpha N_f}{2+\pi\alpha N_f D/8}+\frac{16}{D}\right). \label{Nc2}
\end{equation}
The second term in the right-hand side of Eq.~(\ref{Nc2}), namely $32/(\pi^2 D)=2/(\pi^2 x_0)$, is the nontrivial contribution of the Yukawa action. Indeed, it is a consequence of the second integral in Eq.~(\ref{sigmap3}), which is dependent on $x_0$. Because of scale invariance, the critical point is only dependent on $\alpha$. From Eq.~(\ref{Nc2}), we find $0.81\leq N_c(\alpha)\leq 4.05$ for $D=4$ and $1.61\leq N_c(\alpha)\leq 8.10$ for $D=2$, where the lower and upper limits have been calculated from the $\alpha\rightarrow 0$ and $\alpha\rightarrow\infty$ cases, respectively. For QED3, the critical number is $N^{{\rm QED3}}_c=32/\pi^2\approx 3.24$ \cite{Appelquist1985}. For graphene, the critical point $N_c$ may be modified by a substrate or by renormalization of the Fermi velocity, because $\alpha=e^2/(4\pi \epsilon v_F)$, where $\epsilon$ and $v_F$ are the dielectric constant and Fermi velocity, respectively.

Because $N_f$ is dependent on $\alpha$, it is interesting to obtain a critical fine-structure constant $\alpha_c$. Using Eq.~(\ref{Nc}) in Eq.~(\ref{gamma}), we have
\begin{equation}
\gamma=\frac{1}{2}\sqrt{\frac{2}{\pi^2N_f}\left(\frac{2\lambda}{2+\lambda D/32}+\frac{16}{D}\right)-1}.
\label{Eq49}
\end{equation}
We define a $\lambda_c$ such that for $\lambda<\lambda_c$, the factor $\gamma$ is complex, hence, no mass function exist. From Eq.~(\ref{Eq49}), we obtain that
\begin{equation}
\frac{2}{\pi^2N_f}\left(\frac{2\lambda_c}{2+\lambda_c D/32}+\frac{16}{D}\right)=1.
\label{lamc}
\end{equation}
Solving Eq.~(\ref{lamc}) for $\lambda_c$, we have
\begin{equation}
\lambda_c=\frac{16}{\pi D }\frac{32-DN_f \pi^2}{DN_f\pi^2-160}.  \label{lambdac}
\end{equation}
Furthermore, using $4\pi\alpha_c N_f=\lambda_c$, we find
\begin{equation}
\alpha_c=\frac{16\pi(\frac{32}{\pi^2 DN_f}-1)}{DN_f\pi^2-160}.  \label{alphac}
\end{equation}

Note that $\gamma$
is real for $\alpha \geq \alpha_c$, because the quantity between parentheses in Eq.~(\ref{Eq49})
is monotonically increasing. Eq.~(\ref{alphac}) shows that the critical constant is a function of $D N_f$. It precisely shows that the dynamical mass generation is independent on the parameter $D$. Let us clarify this result. Assume we are in a representation $D=2$ with $N_f=4$, as usually is the case for graphene. $N_f=4$ describes the two spins $\uparrow,\downarrow$ and the two valleys K and K' (internal degrees of freedom). We perform the same calculations in other representation $D=4$, hence decreasing the flavor number to $N_f=2$ (only spins or valleys). It follows that $D N_f=8$ for both cases, therefore, $\alpha_c\approx 0.36$ is obtained independent on the representation. Furthermore, because this result is much less than $\alpha_c= 1.02$ with $G=0$ \cite{PRX2015}, we conclude that the presence of some other microscopic interaction is likely to favor the phase with mass generation, even if this last is weak. 

\section{\textbf{VII. The Physical Mass $m_{\rm{ph}.}$}}

The broken phase and its critical point may be obtained from the function $\Sigma(p)$ in Eq.~(\ref{full1}). This, however, is not the physical mass $m_{\rm{ph}}$, which is expected to be observed in the phase with broken symmetry. Here, we shall calculate the renormalized spectrum, namely, $E_{\pm}(\textbf{p})=\pm\sqrt{\textbf{p}^2+m_{\rm{ph}}^2}$. Note that $2|m_{\rm{ph}}|$ is the energy gap at the Dirac point $\textbf{p}=0$.

By making a Taylor expansion around $m_{\rm{ph}}$, the mass function reads
\begin{eqnarray}
\Sigma(p)=\Sigma(p=m_{\rm{ph}.})+(\gamma^\mu p_\mu-m_{\rm{ph}.})\frac{\partial\Sigma(p)}{\partial p}\huge|_{p=m_{\rm{ph}.}}+...\end{eqnarray}
and imposing
\begin{equation}
\Sigma (p=m_{\rm{ph}.})=m_{\rm{ph}.} , \label{recond}
\end{equation}
we may write the full fermion propagator as
\begin{eqnarray}
S_F(p)&=&\frac{1}{\gamma^\mu p_\mu-\Sigma(p)}\nonumber\\
&=&\frac{1}{(\gamma^\mu p_\mu-m_{\rm{ph}.})(1-\frac{\partial\Sigma(p)}{\partial p}\huge|_{p=m_{\rm{ph}.}}+...)}
\nonumber\\
&=&\frac{\gamma^\mu p_\mu+m_{\rm{ph}.}}{(p^2 - m_{\rm{ph}.}^2)(1-\frac{\partial\Sigma(p)}{\partial p}\huge|_{p=m_{\rm{ph}.}}+...)}.
\label{sfsup}
\end{eqnarray}

We see that  $m_{\rm{ph}.}$ is the pole of the full electron propagator at zero momentum, hence, the physical mass. Note that the condition for calculating this is given by Eq.~(\ref{recond}), which applies only the real part of the self-energy in Eq.~(\ref{Rewritten}). Using $g\Sigma(p)=f(gp)$ (for the real parts), we have
\begin{equation}
g \Sigma(p)=\frac{F}{\sqrt{gp}}\cos\left[\gamma \ln\left(gp\right)+\theta_0\right]. \label{realsig}
\end{equation}

Next, we use Eq.~(\ref{recond}) to calculate $m_{\rm{ph}}$. We choose $\theta_0=0$ and $F=1$, without loss of generality. Therefore, using Eq.~(\ref{realsig}) at $p=m_{\rm{ph} }$ and defining $-z\equiv \gamma \ln (g m_{\rm{ph}})$, we have
\begin{equation}
m_{\rm{ph}}= g^{-1} \exp\left(-\frac{z}{\gamma}\right),
\end{equation}
where $z$ are solutions of the transcendental equation
\begin{equation}
\exp\left(-\frac{3z}{2\gamma}\right)= \cos z.  \label{zeq}
\end{equation}

The number of dynamically generated masses is the same as in Ref.~\cite{PRX2015}, but in our case $g^{-1}$ is the natural cutoff for the theory. For $\alpha\rightarrow \alpha_c$ and $\gamma\approx 0$, Eq.~(\ref{zeq}) gives $z\rightarrow z_n=(2n+1)\pi/2$ with $n$ integer. In comparison with the results in Ref.~\cite{PRX2015} with $g=0$, we conclude that microscopic interactions are not likely to cancel the quantum valley Hall effect, because the quantization in the energy levels survives even with $g\neq 0$. This is not surprisingly in the view of topological insulator theory, because it is well known that the quantum Hall effect experimentally occur in the presence of impurities or electron-phonon interactions. Here, we have the very same conclusion, but for the interaction-driven quantum valley Hall effect in Ref.~\cite{PRX2015}.

\section{ \textbf{VIII.\, Summary and Outlook}}

We have described the dynamical symmetry breaking in a two-dimensional system, consisting of massless Dirac fermions with two types of interactions. One is the electromagnetic, described by the PQED approach and the other is, essentially, a Yukawa action, which originates from some microscopic physical effect, such as electron-phonon interaction. This seems to be case for honeycomb systems at low energies, see \cite{Saito} for the case of graphene. 

We consider the zero temperature case at both SU(2) and SU(4) representations, where parity and chiral symmetries are broken in the massive phase, respectively. We show that the criticality of the theory is independent on the specific representation $D$, when we consistently consider the changing on the flavor number $N_f$. It is shown that for $D N_f=8$, the critical coupling constant is $\alpha_c\approx 0.36$. Hence, the presence of some microscopic interaction is likely to improve the possibility of generating a mass in the electronic spectrum. The mass function of the matter field has been calculated from the Schwinger-Dyson equations, using the large$-N_f$ expansion for both the Gauge and scalar fields. This is called unquenched-rainbow approximation in literature. Numerical results show that our analytical approximations are in good agreement with the full integral equation. Furthermore, we have shown that Yukawa interaction does not change the quantized feature of the energy levels, which have been calculated in Ref.~\cite{PRX2015} for the case of pure PQED.

It is a experimental challenge to generate a finite mass for massless Dirac particles in graphene. From the theoretical view, the task is to find a lesser critical coupling constant $\alpha_c$, such that the electromagnetic interaction would be enough to generate a finite mass. Having in mind that, in two-dimensional materials, there exist several microscopic interactions, beyond the Coulomb repulsion. Hence, we believe our results are a 
encouraging step for deriving symmetry broken phases. Nevertheless, we have provided a analytical result for the mass function of a nonlocal model interacting with a four-fermion action. Several generalizations of this paper would be investigated, for instance, the role of an external magnetic field, the Fermi velocity, chemical potential, and finite temperature. Furthermore, we may generalize our four-fermion interaction to a Thirring version or other important microscopic interaction. We shall study these cases elsewhere. 

\section{\textbf{acknowledgments}}

This work was supported in part by CNPq (Brazil), CAPES (Brazil), and FAPERJ (Brazil).

\textbf{\section{\textbf{Appendix A: Wave Function Renormalization}}}

\begin{figure}[htb]
\centering
\includegraphics[scale=1.2]{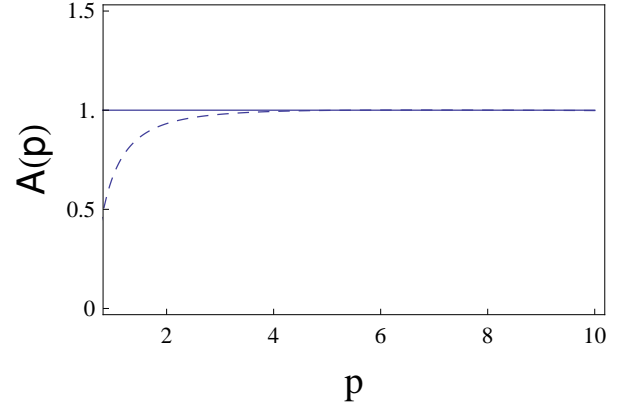}
\caption{(Color online) Comparison between the numerical results of Eq.~(\ref{ap}) and the analytical solution $A(p)= 1.0$. The dashed line is the numerical solution with $\Sigma(p)=0$ (symmetric phase). We choose $\alpha=2.2$, $D=4$, $N_f=10$, $g=0.1$, and $\Lambda=10$ to perform the numerical calculations. The analytical approximation describes well the numerical result, in particular, for large momentum. } \label{fig3}
\end{figure}

In this appendix, we derive the integral equation for the wave function renormalization $A(p)$, defined in Eq.~(\ref{full1}). In order to do so, we multiply Eq.~(\ref{sd}) by $\gamma^{\mu}p_{\mu}$. Thereafter, we calculate the traces over the Dirac matrices. It yields
\begin{equation}
A(p) = 1 + \frac{1}{N_f D p^2}{\rm Tr}\left[\Xi(p)\gamma^{\alpha}p_{\alpha}\right]. \label{apwavefunction1}
\end{equation}
Using Eq.~(\ref{sea}), Eq.~(\ref{seg}), and Eq.~(\ref{full1}) into Eq.~(\ref{apwavefunction1}), after calculating the trace over Dirac matrices, we find
\begin{eqnarray}
A(p) &=& 1 + \frac{2\lambda}{N_f p^2} \int\frac{d^{3}k}{(2\pi)^{3}}\frac{A(k) G(q)}{A(k)^{2}k^{2}+\Sigma^{2}(p)}\frac{(k \cdot q)(p \cdot q)}{q^2} + \nonumber \\
&+& \frac{1}{N_{f}}\int\frac{d^{3}k}{(2\pi)^{3}}\frac{p \cdot k}{p^{2}}\frac{A(k) \Delta_{\varphi}(p-k)}{A^{2}(k)k^{2}+\Sigma^{2}(k)}, \label{apwavefunction2}
\end{eqnarray}
where $G(q)$ and $q$ are given by
\begin{equation}
q = p-k, \label{momentumq}
\end{equation}
and
\begin{equation}
G(q) = \frac{1}{\sqrt{q^2}\left(2+\frac{\lambda D}{32}\right)}. \label{propagatephoton}
\end{equation}
We can exaclty solve angular integral the renormalization function to obtain

\begin{eqnarray}
A(p) &=& 1-\frac{B_{1}}{p^{3}} \int_{0}^{\infty}dk \frac{k A(k)}{A^{2}(k)k^{2}+\Sigma^{2}(p)} \mathcal{B}_{1}(k,p) + \nonumber \\ 
&-&\frac{B_{2}}{p^{3}} \int_{0}^{\infty}dk \frac{k A(k)}{A^{2}(k)k^{2}+\Sigma^{2}(p)} \mathcal{B}_{2}(k,p)  \label{ap}
\end{eqnarray}
where,
\begin{equation}
B_{1}= \frac{2\lambda}{N_{f}}\frac{1}{16\pi^{2}}\frac{1}{\left(2+\frac{\lambda D}{32}\right)},
\end{equation}

\begin{equation}
B_{2} = \frac{1}{DN_{f}} \frac{1}{3\pi^{3}},
\end{equation}

\begin{eqnarray}
\mathcal{B}_{1}(k,p) = \frac{1}{3}(p+k)^{3}-\frac{1}{3}|p-k|^{3} + (p^{2}-k^{2}) \times\nonumber \\
\times\left[\, (p-k) \operatorname{Sgn}(p+k) - (p+k)\operatorname{Sgn}(p-k)\, \right],
\end{eqnarray}
and
\begin{eqnarray}
\mathcal{B}_{2}(k,p) &=&\left[4k^{2}+4p^{2}-6(x_{0}g)^{-2}\right]\left(\, |p-k|-|p+k|\, \right) + \nonumber \\ &+& 4kp \left[|p-k|+|p+k|-3(x_{0}g)^{-1}\right] + \nonumber \\ &-& 6(x_{0}g)^{-1}\left[k^{2}+p^{2}-(x_{0}g)^{-2}\right] \times \nonumber\\ &\times& \ln\left[\frac{(x_{0}g)^{-1}+|p-k|}{(x_{0}g)^{-1}+|p+k|}\right].
\end{eqnarray}
In Fig.~\ref{fig3} , we compare the numerical results of this integral equation with the analytical solution $A(p)=1$. As expected, a very good agreement is found, specially, for large momentum, which the integral equation may be solved.

 \section{\textbf{Appendix B: The Analytical Approach}}

In order to convert Eq.~(\ref{eqint1}) into a differential equation, it is convenient to perform an approximation in the logarithmic kernel. In the lowest order, we have
\begin{eqnarray}
\ln\left(\frac{x_0^{-1}+|x+y|}{x_0^{-1}+|x-y|}\right)&\approx& \frac{2 y}{x+x_0^{-1}}\Theta(x-y)\nonumber\\
&+&\frac{2 x}{y+x_0^{-1}}\Theta(y-x).
\end{eqnarray}

Introducing a ultraviolet cutoff $\Lambda$, we find
\begin{eqnarray}
f(x)&=&\frac{2 C_1}{x}\left[\int_0^x \frac{y^2 f(y) dy}{y^2+f^2(y)}+ x \int_x^{g \Lambda} \frac{y f(y) dy}{y^2+f^2(y)} \right]\nonumber\\
&+&g C_2[\int_0^x \frac{y^2 f(y) dy}{y^2+f^2(y)}\frac{2}{x(x+x^{-1}_0)}\nonumber \\
&+&\int_x^{g\Lambda} \frac{y f(y) dy}{y^2+f^2(y)}\frac{2}{(y+x^{-1}_0)}]. \label{apefx}
\end{eqnarray}
In general grounds, the derivative of an arbitrary function $F(p)$, given by
\begin{equation}
F(p)=\int_{v(p)}^{u(p)} dk {\cal K}(k,p)
\end{equation}
is
\begin{equation}
\frac{dF(p)}{dp}=\int_{v(p)}^{u(p)} dk \frac{\partial {\cal K}(k,p)}{\partial p}+ \frac{\partial u}{\partial p} {\cal K}\textbf{(}p, u(p)\textbf{)}-\frac{\partial v}{\partial p} {\cal K}\textbf{(}p, v(p)\textbf{)}, \label{c5}
\end{equation}
where $v(p)$ and $u(p)$ are also arbitrary functions. Using this in Eq.~(\ref{apefx}), we find
\begin{eqnarray}
\frac{df}{dx}&=&-\frac{2C_1}{x^2} \int_0^x \frac{y^2 f(y) dy}{y^2+f^2(y)}\nonumber\\
&+&g C_2 \frac{d}{dx}\left[\frac{2}{x(x+x^{-1}_0)}\right] \int_0^x \frac{y^2 f(y) dy}{y^2+f^2(y)}. \label{apefx2}
\end{eqnarray}

For practical reasons, we define
\begin{equation}
h^{-1}(x)=1-\frac{g C_2 x^2}{2C_1}\frac{d}{dx}\left[\frac{2}{x(x+x^{-1}_0)}\right].
\end{equation}

By deriving Eq.~(\ref{apefx2}), we find
\begin{equation}
\frac{d}{dx}\left[ h(x) x^2 \frac{df}{dx}\right]+\frac{N_c}{4N_f}\frac{x^2 f(x)}{x^2+f^2(x)}=0,
\end{equation}
where $N_c$ is given by Eq.~(\ref{Nc}). Finally, we consider the asymptotic limits $x\gg f(x)$, hence $h(x)\approx 1$. In this case, we obtain Eq.~(\ref{eulereq}). 

\section{\textbf{Appendix C: Comparison Between Numerical and Analytical Results}}

\begin{figure}[htb]
\centering
\includegraphics[scale=1.0]{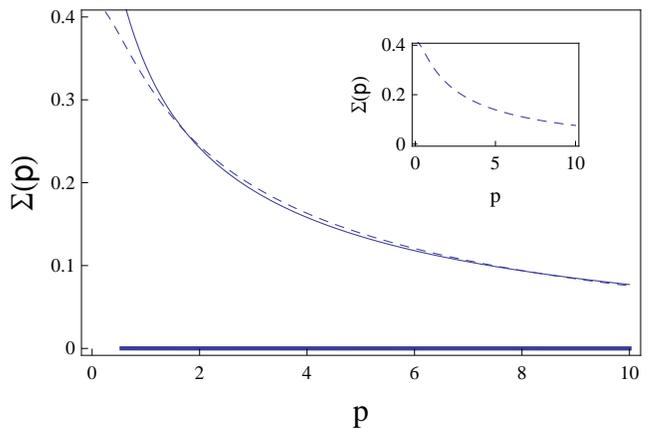}
\caption{(Color online) Comparison between the numerical results of Eq.~(\ref{eqint0}) and the analytical solution in Eq.~(\ref{Rewritten}). The common line is the analytical solution, the dashed line is the numerical result with $\alpha=2.2$, and the thick line is the numerical result with $\alpha=0.3$. We choose the following parameters: $\Lambda=10$, $g=0.1$, $D=4$, and $N_f=2.0$ for all the lines. The critical coupling constant is $\alpha_c=0.36$, given by Eq.~(\ref{alphac}). For the analytical solution, we consider $A_++A_-=2.7/350$ and $A_+-A_-=i2.7/350$, the best fitting parameters. The inset is the numerical solution for $\alpha=2.2$ to facilitate its visualization. } \label{fig1}
\end{figure}

In this appendix, we perform some numerical tests to verify the validity of the analytical approaches adopted in this paper. In order to obtain the numerical results for the mass function, we consider the full integral equation for $\Sigma(p)$ with $A(p)=1$, given by Eq.~(\ref{eqint0}). The numerical result in Fig.~\ref{fig1} is obtained after we convert the momentum-dependent kernel in  Eq.~(\ref{eqint0}) into a system of nonlinear algebraic equations, using the repeated trapezoidal quadrature rule. Furthermore, it is mandatory to include a cutoff $\Lambda$ to perform numerical calculations. Without loss of generality, we take $\Lambda=10$ (in units of energy). For more details about these steps, see Ref.~\cite{VLWJF} which has this procedure for PQED at zero temperature and Ref.~\cite{CSBTemperature} at finite temperature.

The analytical solution $\Sigma(p)$ is promptly obtained from Eq.~(\ref{Rewritten}). Indeed, it has been shown that $\Sigma(p)=g^{-1}f(gp)$ in Sec.~VI. After we assume $A_++A_-=2.7/350$ and $A_+-A_-=i2.7/350$, a very good agreement is observed between numerical and analytical results, see Fig.~\ref{fig1}. This assumption is possible because $A_+$ and $A_-$ are arbitrary constants as we discussed in Sec.~VI. Eq.~(\ref{alphac}) yields the critical point. Using $D=4$ and $N_f=2.0$, we have $\alpha_c\approx 0.36$. The numerical results are also in agreement with the fact that dynamical mass generation only occurs for $\alpha\geq \alpha_c$.

\end{document}